\documentclass[pra,aps,showpacs,onecolumn,twoside,superscriptaddress]{revtex4}

\usepackage{amsmath,amsfonts,amssymb,color,epsfig,graphics,graphicx,latexsym,revsymb,theorem,url,verbatim}

\newtheorem{definition}{Definition}
\newtheorem{proposition}[definition]{Proposition}
\newtheorem{lemma}[definition]{Lemma}

\newtheorem{theorem}[definition]{Theorem}
\newtheorem{corollary}[definition]{Corollary}
\newtheorem{conjecture}[definition]{Conjecture}

\newtheorem{remark}[definition]{Remark}
\newtheorem{example}[definition]{Example}

\def\squareforqed{\hbox{\rlap{$\sqcap$}$\sqcup$}}
\def\qed{\ifmmode\squareforqed\else{\unskip\nobreak\hfil
\penalty50\hskip1em\null\nobreak\hfil\squareforqed
\parfillskip=0pt\finalhyphendemerits=0\endgraf}\fi}
\def\endenv{\ifmmode\;\else{\unskip\nobreak\hfil
\penalty50\hskip1em\null\nobreak\hfil\;
\parfillskip=0pt\finalhyphendemerits=0\endgraf}\fi}
\newenvironment{proof}{\noindent \textbf{{Proof.~} }}{\qed}
\def\Dbar{\leavevmode\lower.6ex\hbox to 0pt
{\hskip-.23ex\accent"16\hss}D}
\makeatletter
\def\url@leostyle{%
  \@ifundefined{selectfont}{\def\UrlFont{\sf}}{\def\UrlFont{\small\ttfamily}}}
\makeatother
\def\bcj{\begin{conjecture}}
\def\ecj{\end{conjecture}}
\def\bcr{\begin{corollary}}
\def\ecr{\end{corollary}}
\def\bd{\begin{definition}}
\def\ed{\end{definition}}
\def\bea{\begin{eqnarray}}
\def\eea{\end{eqnarray}}
\def\bem{\begin{enumerate}}
\def\eem{\end{enumerate}}
\def\bex{\begin{example}}
\def\eex{\end{example}}
\def\bim{\begin{itemize}}
\def\eim{\end{itemize}}
\def\bl{\begin{lemma}}
\def\el{\end{lemma}}
\def\bpf{\begin{proof}}
\def\epf{\end{proof}}
\def\bpp{\begin{proposition}}
\def\epp{\end{proposition}}
\def\br{\begin{remark}}
\def\er{\end{remark}}
\def\bt{\begin{theorem}}
\def\et{\end{theorem}}

\newcommand{\nc}{\newcommand}

\def\a{\alpha}
\def\b{\beta}

\def\e{\epsilon}

\def\k{\kappa}
\def\l{\lambda}
\def\m{\mu}

\def\p{\pi}
\def\r{\rho}
\def\s{\sigma}

\def\G{\Gamma}
\def\D{\Delta}

\nc{\bbC}{{\mathbb{C}}}

\nc{\cA}{{\cal A}} \nc{\cB}{{\cal B}} \nc{\cC}{{\cal C}}
\nc{\cD}{{\cal D}} \nc{\cE}{{\cal E}} \nc{\cF}{{\cal F}}
\nc{\cG}{{\cal G}} \nc{\cH}{{\cal H}} \nc{\cI}{{\cal I}}
\nc{\cJ}{{\cal J}} \nc{\cK}{{\cal K}} \nc{\cL}{{\cal L}}
\nc{\cM}{{\cal M}} \nc{\cN}{{\cal N}} \nc{\cO}{{\cal O}}
\nc{\cP}{{\cal P}} \nc{\cR}{{\cal R}} \nc{\cS}{{\cal S}}
\nc{\cT}{{\cal T}} \nc{\cU}{{\cal U}} \nc{\cV}{{\cal V}}
\nc{\cW}{{\cal W}} \nc{\cX}{{\cal X}} \nc{\cZ}{{\cal Z}}

\nc{\hA}{{\hat{A}}} \nc{\hB}{{\hat{B}}} \nc{\hC}{{\hat{C}}}
\nc{\hD}{{\hat{D}}} \nc{\hE}{{\hat{E}}} \nc{\hF}{{\hat{F}}}
\nc{\hG}{{\hat{G}}} \nc{\hH}{{\hat{H}}} \nc{\hI}{{\hat{I}}}
\nc{\hJ}{{\hat{J}}} \nc{\hK}{{\hat{K}}} \nc{\hL}{{\hat{L}}}
\nc{\hM}{{\hat{M}}} \nc{\hN}{{\hat{N}}} \nc{\hO}{{\hat{O}}}
\nc{\hP}{{\hat{P}}} \nc{\hR}{{\hat{R}}} \nc{\hS}{{\hat{S}}}
\nc{\hT}{{\hat{T}}} \nc{\hU}{{\hat{U}}} \nc{\hV}{{\hat{V}}}
\nc{\hW}{{\hat{W}}} \nc{\hX}{{\hat{X}}} \nc{\hZ}{{\hat{Z}}}

\def\diag{\mathop{\rm diag}}
\def\dim{\mathop{\rm Dim}}

\def\rank{\mathop{\rm rank}}

\def\tr{\mathop{\rm Tr}}

\def\GL{{\mbox{\rm GL}}}

\def\dg{\dagger}

\def\ox{\otimes}

\def\sue{\subseteq}

\newcommand{\bra}[1]{\langle#1|}
\newcommand{\ket}[1]{|#1\rangle}
\newcommand{\proj}[1]{| #1\rangle\!\langle #1 |}
\newcommand{\ketbra}[2]{|#1\rangle\!\langle#2|}

\newcommand{\jmp}{J. Math. Phys.}
\newcommand{\jpa}{J. Phys. A}

\newcommand{\pla}{Phys. Lett. A}

\def\bR{{\mbox{\bf R}}}

\def\bC{{\mbox{\bf C}}}

\begin{document}
\title{ Equivalence classes and canonical forms for two-qutrit entangled states of rank four having positive partial transpose}

\author{Lin Chen}
\affiliation{Department of Pure Mathematics and Institute for
Quantum Computing, University of Waterloo, Waterloo, Ontario, N2L
3G1, Canada} \affiliation{Centre for Quantum Technologies, National
University of Singapore, 3 Science Drive 2, Singapore 117542}
\email{cqtcl@nus.edu.sg (Corresponding~Author)}

\def\Dbar{\leavevmode\lower.6ex\hbox to 0pt
{\hskip-.23ex\accent"16\hss}D}
\author {{ Dragomir {\v{Z} \Dbar}okovi{\'c}}}

\affiliation{Department of Pure Mathematics and Institute for
Quantum Computing, University of Waterloo, Waterloo, Ontario, N2L
3G1, Canada} \email{djokovic@uwaterloo.ca}

\begin{abstract}
Let $\cE'$ denote the set of non-normalized two-qutrit entangled 
states of rank four having positive partial transpose (PPT). 
We show that the set of SLOCC equivalence classes of states 
in $\cE'$, equipped with the quotient topology, is
homeomorphic to the quotient $R/A_5$ of the open rectangular
box $R\subset\bR^4$ by an action of the alternating group $A_5$.
We construct an explicit map $\omega:\Omega\to\cE'$, where 
$\Omega$ is the open positive orthant in $\bR^4$, whose image 
$\omega(\Omega)$ meets 
every SLOCC equivalence class $E\subseteq\cE'$. 
Although the intersection $\omega(\Omega)\cap E$ is not necessarily a singleton set, it is always a finite set of  
cardinality at most 60. By abuse of language, we say that  
any state in $\omega(\Omega)\cap E$ is a canonical form of 
any $\r\in E$.
In particular, we show that all checkerboard PPT
entangled states can be parametrized up to SLOCC equivalence by
only two real parameters. We also summarize the known results on
two-qutrit extreme PPT states and edge states, and examine which
other interesting properties they may have. Thus we find the
first examples of extreme PPT states whose rank is different from
the rank of its partial transpose.
\end{abstract}

\date{ \today }


%03.67.Mn Entanglement production, characterization and manipulation
%03.65.Ud Entanglement and quantum nonlocality
%(e.g. EPR paradox, Bell's inequalities, GHZ states, etc.)
%(for entanglement production in quantum information, see 03.67.Mn;
%for entanglement in Bose-Einstein condensates, see 03.75.Gg)
%03.67.-a   Quantum information

%02.20.-a   Group theory
%(for algebraic methods in quantum mechanics, see 03.65.Fd;
%for symmetries in elementary particle physics, see 11.30.-j)
%03.65.Ta Foundations of quantum mechanics; measurement theory
%(for optical tests of quantum theory, see 42.50.Xa)
%03.65.Wj State reconstruction, quantum tomography
%03.67.Dd   Quantum cryptography
%03.67.Hk   Quantum communication
%05.30.Jp   Boson systems
%02.30.Hq   Ordinary differential equations
%02.30.Nw   Fourier analysis

\maketitle

%\tableofcontents

%\vspace{2pc} \noindent{\it Keywords}: multipartite entanglement,
%additivity, relative entropy of entanglement, geometric measure of
%entanglement, global robustness, symmetric state, antisymmetric
%state

%\section{\label{sec:3x3rankfour} clues on $3\times3$ states of rank four}
%
%
%
%For convenience, we list some known results on distillability under
%LOCC for $M\times N$ states.
%
% \bpp
% \label{bm:1distillability}
% We have the following result on 1-distillability of bipartite NPT states.
% \bem
%\item
%  $2\times N$ NPT states are 1-distillable \cite{bbp96,hhh97prl,dss00}.
%\item
%  NPT states of rank at most three are 1-distillable \cite{cc08,hst99}.
%\item
%  The states violating the reduction criterion (e.g. the isotropic states) are 1-distillable \cite{hh99}.
%\item
%  The $M\times N$ states of rank less than or equal to $N$ are 1-distillable \cite{hst99,cd11}.
% \eem
% \epp
%
%In this section we study the 1-distillability of $3\times3$ NPT
%states $\r$ with rank four. We start by giving a few clues.
%
% \bem
%\item
% The states violating Proposition \ref{bm:1distillability} is
% 1-distillable. Can we figure out the expression of such states $\r$
% ?
%\item
% We should study the Bures-Peres family of states which is just a state $\r$.
%\item
% Can we locally project the example by Waterous onto a 1-non-distillable
% state $\r$ ?
% \eem

\section{\label{sec:introduction} Introduction}

Entanglement plays the essential role in quantum information
processing, such as quantum teleportation \cite{bbc93}, computing
\cite{nc00} and cryptography \cite{ekert91}. Entanglement also
reveals a fundamental difference between the quantum and
classical world which may be detected by Bell inequallity and
other means \cite{bell64}. In spite of these striking facts,
it is a hard problem to decide whether a given quantum state
is entangled \cite{gurvits03}.

A non-entangled state, also known as a separable state, is by
definition a convex sum of product states \cite{werner89}. For a
bipartite state $\r$ acting on the Hilbert space $\cH:=\cH_A \ox
\cH_B$, the partial transpose computed in an orthonormal (o.n.)
basis $\{\ket{a_i}\}$ of system A, is defined by
$\r^\G=\sum_{ij}\ketbra{a_j}{a_i}\ox\bra{a_i}\r\ket{a_j}$. The
dimensions of $\cH_A$ and $\cH_B$ are denoted by $M$ and $N$,
respectively. (We assume that $M,N\ge2$.) We say that $\r$ is a
$k\times l$ {\em state} if its local ranks are $k$ and $l$, i.e.,
$\rank\r_A=k$ and $\rank\r_B=l$.
We say that $\r$ is a PPT [NPT] state if $\r^\G\ge0$ [$\r^\G$ has
at least one negative eigenvalue]. Evidently, a separable state
must be PPT. The converse is true only if $MN\le6$ \cite{peres96,hhh96}. The first examples of two-qutrit PPT entangled
states (PPTES) were constructed (in purely mathematical context)
by Choi and St{\o}rmer in the 1980s \cite{choi82,stormer82}.
They were introduced in 1997 \cite{horodecki97} into quantum
information theory. The Choi's example has been generalized in \cite{hkp03}. The PPTES can be used for many tasks, e.g. the
extraction of distillable key \cite{hhh05}.

A systematic method of constructing two-qutrit PPTES $\r$ of
rank four was proposed in 1999 \cite{bdm99} by using the unextendible product bases (UPB).
This construction is indeed universal: it was proved in 2011
\cite{cd11JMP,skowronek11} that any such $\r$ can be constructed
by using a UPB.
Any such $\r$ can be converted by stochastic local 
operations and classical communications (SLOCC) into a canonical form depending on for positive parameters, see 
Eq. \eqref{ea:Blokovi} and the subsequent paragraph. 
This canonical form is a minor modification of the one 
constructed in \cite[Eq. (108)]{cd11JMP}. 
We demonstrate our result by using the well-known Pyramid and
Tiles UPB \cite{DiV03}.
The set of SLOCC equivalence classes of two-qutrit PPTES
carries the natural quotient topology. We show that this
topological space is homeomorphic to the quotient of the open
rectangular box $R\subset\bR^4$ by an action of the alternating
group $A_5$.
The checkerboard family of PPTES is an early well-known example of an infinite family of two-qutrit PPTES of rank four
\cite{BrPeres}.
We show that, under SLOCC, the checkerboard PPTES can be parametrized by only two real parameters.
These results clarify the structure of $3\times 3$ PPTES of
rank four.

We say that a PPT state $\r$ is {\em extreme} if it is an
extreme point of the compact convex set of all normalized PPT
states. Every PPT state is a convex linear combination of
extreme states. For convenience we say that a non-normalized state is {\em extreme} if its normalization is extreme.

We say that a PPT state $\r$ is an {\em edge state} if there
is no product vector $\ket{a,b}\in\cR(\r)$ such that
$\ket{a^*,b}\in\cR(\r^\G)$. Thus an edge state must be a PPTES.
It is easy to see that any entangled extreme state is an edge
state, but the converse is false. It has been proved that any
$3\times3$ PPTES of rank four is an extreme state
\cite{cd11JMP}. They are PPTES of lowest rank. Both extreme and
edge states in higher dimensions have been extensively studied
in recent years \cite{cd11JMP,skowronek11,cd12,ko12}. Among
them, $3\times3$ extreme and edge states have the simplest structure. We will give a summary of the recent progress on
these states, and point out some unknown cases.
We explicitly construct states for these cases, including
extreme PPTES of birank $(5,6)$ and $(5,7)$, as well as
non-extreme edge states of birank $(6,6)$, $(5,6)$ and $(5,7)$.
(We refer to the ordered pair $(\rank\r,\rank\r^\G)$ as the
{\em birank} of $\r$.)

The content of our paper is as follows. In Sec.
\ref{sec:preliminary} we provide the technical tools used in the
paper. In Sec. \ref{sec:equivalence} we consider the set $\cE'$
of $3\times3$ PPTES of rank four, which is invariant under the
natural action of the group $G$ of local invertible
transformations. In Theorem \ref{thm:ParPPTES} we prove that any 
$\r\in\cE'$ can be converted under SLOCC
into the canonical form. We also prove, see Theorem \ref{Klasif}, 
that the quotient spaces $\cE'/G$ and $R/A_5$ mentioned in the 
abstract are homeomorphic.
In Sec. \ref{sec:checkerboard}, we construct a family of 
checkerboard PPTES depending on two positive 
parameters $u$ and $v$, see Proposition \ref{KanCheckerBoard}.
We show that any checkerboard PPTES is SLOCC equivalent to 
a member of this family. By using invariants, we find a simple 
characterization of checkerboard PPTES (see Proposition 
\ref{pp:karakterizacija}).
In Sec. \ref{sec:extreme} we summarize the recent progress on
$3\times 3$ extreme and edge states.
We examine some of the known states of this type to find out what
other properties they may have.
E.g. we observe that there exist extreme states of birank $(r,s)$
with $r\ne s$. Apparently, this observation is new.
The results are presented in Table \textbf{I} at the end of this section. We also propose several open problems.

\section{\label{sec:preliminary} Preliminaries}

We shall write $I_k$ for the identity $k\times k$ matrix.
We denote by $\cR(\r)$ and $\ker \r$ the range and kernel of a
linear map $\r$, respectively. From now on, unless stated
otherwise, the states will not be normalized. We shall denote by
$\{\ket{i}_A:i=0,\ldots,M-1\}$ and
$\{\ket{j}_B:j=0,\ldots,N-1\}$ o.n. bases of $\cH_A$ and
$\cH_B$, respectively.
The subscripts A and B will be often omitted.
Any state $\r$ of rank $r$ can be represented as
 \bea
 \label{ea:MxN-State}
\r=\sum_{i,j=0}^{M-1} \ketbra{i}{j}\ox C_i^\dag C_j,
 \eea
where the $C_i$ are $R\times N$ matrices and $R$ is an arbitrary
integer $\ge r$. In particular, one can take $R=r$.
We shall often consider $\r$ as a block matrix
$\r=C^\dag C=[C_i^\dag C_j]$, where $C=[C_0~C_1~\cdots~C_{M-1}]$
is an $R\times MN$ matrix. Thus $C_i^\dag C_j$ is the matrix of
the linear operator $\bra{i}_A \r\ket{j}_A$ acting on $\cH_B$.

In physics, the density matrix $\r$ describes the systems $AB$. In
many cases we only need to describe one system by the reduced
density matrices $\r_A=\tr_B \r$ and $\r_B=\tr_A\r$. For these
matrices, using Eq. \eqref{ea:MxN-State} we have the formulae
 \bea \label{RedStates}
\r_B=\sum_{i=0}^{M-1} C_i^\dag C_i; \quad \r_A=[\tr C_i^\dag C_j],
\quad i,j=0,\ldots,M-1.
 \eea

It is easy to verify that the range of $\r$ is the column space
of the matrix $C^\dag$ and that
 \bea \label{JezgroRo}
\ker\r=\left\{\sum_{i=0}^{M-1}\ket{i}\ox\ket{y_i}:
\sum_{i=0}^{M-1}C_i\ket{y_i}=0\right\}.
 \eea
In particular, if $C_i\ket{j}=0$ for some $i$ and $j$ then
$\ket{i,j}\in\ker\r$.

For any bipartite state $\r$ we have
 \bea
 \label{ea:rhoBGamma=rhoB}
\left( \r^\G \right)_B &=& {\tr}_A \left( \r^\G \right) =
{\tr}_A \r = \r_B, \\
 \label{ea:rhoAGamma=rhoGammaA}
\left( \r^\G \right)_A &=& {\tr}_B \left( \r^\G \right) = \left(
{\tr}_B \r \right)^T = ( \r_A )^T.
\eea
(The exponent T denotes transposition.) Consequently,
 \bea
\rank \left( \r^\G \right)_{A,B} = \rank \r_{A,B}.
 \eea
If $\r$ is an $M\times N$ PPT state, then $\r^\G$ is too.
If $\r$ is a PPTES so is $\r^\G$, but they may have different
ranks.

Let us now recall some basic results from quantum information
regarding the separability and PPT properties of bipartite
states. Let us start with the basic definition.

 \bd
We say that two bipartite states $\r$ and $\s$ are {\em
equivalent under stochastic local operations and classical communications} ({\em SLOCC-equivalent}, or just {\em equivalent}) if there exists an invertible local operator
(ILO) $V=V_A\ox V_B$ such that $V\r V^\dag=\s$ \cite{dvc00}.
 \ed

This is the physical realization of ILO. Thus the equivalence classes of states are just the orbits under
the action of the group $G:=\GL_3(\bC)\times\GL_3(\bC)$.
It is easy to see that any ILO transforms PPT, entangled, or
separable state into the same kind of state. We shall often use ILOs to simplify the density matrices of states.

Let us recall from \cite[Theorem 22]{cd11JPA} and
\cite[Theorems 17,22]{cd11JMP} the main facts about the
$3\times3$ PPT states of rank four. Let $M=N=3$ and let $\cU$ denote the set of unextendible product bases in $\cH$.
For $\{\psi\}\in\cU$ we denote by $\Pi\{\psi\}$ the normalized
state $(1/4)P$, where $P$ is the orthogonal projector onto
$\{\psi\}^\perp$.
We say that a subspace of $\cH$ is {\em completely entangled}
(CES) if it contains no product vectors.
(We require product vectors to be nonzero.)
For counting purposes we do not distinguish product vectors
which are scalar  multiples of each other.

We give a formal definition of the term
``general position'' \cite[Definition 7]{cd12}.

 \bd \label{def:GenPos}
We say that a family of product vectors
$\{\ket{\psi_i}=\ket{\phi_i}\ox\ket{\chi_i}:i\in I\}$ is in
{\em general position} (in $\cH$) if for any $J\sue I$ with
$|J|\le M$ the vectors $\ket{\phi_j}$, $j\in J$, are linearly
independent and for any $K\sue I$ with $|K|\le N$ the vectors
$\ket{\chi_k}$, $k\in K$, are linearly independent.
 \ed

\bt \label{thm:3x3PPTstates} $(M=N=3)$ For a $3\times3$ PPT
state $\r$ of rank four, the following assertions hold.

(i) $\r$ is entangled if and only if $\cR(\r)$ is a CES.

(ii) If $\r$ is separable, then it is either the sum of four
pure product states or the sum of a pure product state and a
$2\times2$ separable state of rank three.

(iii) If $\r$ is entangled, then

\quad (a) $\r$ is extreme;

\quad (b) $\rank \r^\G=4$;

\quad (c) $\r=V~\Pi\{\psi\}~V^\dag$ for some $V\in G$ and some
$\{\psi\}\in\cU$;

\quad (d) $\ker\r$ contains exactly 6 product vectors, and these
vectors are in general position. \et

The first assertion of the following proposition is \cite[Theorem
23]{cd11JMP}.

 \bpp
 \label{prop:3x3rank4PPTES,invariantexpression}
 $(M=N=3)$ Any $\r\in\cE'$ is SLOCC equivalent
to one which is invariant under partial transpose, i.e., there
exist $A,B\in\GL_3(\bC)$ such that $\s:=A\ox B~\r~A^\dg\ox B^\dg$ 
satisfies the equality $\s^\G=\s$. Moreover, we may assume that 
$\s=C^\dag C$ where $C=[C_0~C_1~C_2]$ and
 \bea \label{ea:Blokovi}
 C_0= \left[\begin{array}{ccc}
0 & a & b \\
0 & 0 & 1 \\
0 & 0 & 0 \\
0 & 0 & 0
\end{array}\right],\quad
 C_1= \left[\begin{array}{ccc}
0 & 0 & 0 \\
0 & 0 & c \\
0 & 0 & 1 \\
1 & 0 & -1/d
\end{array}\right],\quad
 C_2= \left[\begin{array}{ccc}
0 & -1/b & 0  \\
0 & 1 & 0  \\
1 & -c & 0 \\
d & 0 & 0
\end{array}\right]; \quad a,b,c,d>0.
\eea
 \epp

The weaker form of the second assertion, with $a$ being just
real and nonzero, follows from the proof of \cite[Theorem 23]{cd11JMP} and \cite{cd11JMPe}.
The stronger claim that (like $b,c,d$) $a$ can also
be chosen to be positive will be proved in Theorem
\ref{thm:ParPPTES}.
For convenience, we say that any state $\s=C^\dag C$, where  
$C$ is defined as in this proposition with $a,b,c,d>0$, is in 
the {\em canonical form}. 

As an application, let us recall the following result \cite[Theorem
24]{cd11JMP}.

 \bpp \label{prop:3x3rank4PPTES,samerange}
 $(M=N=3)$ If the normalized states $\r$ and $\r'$ are
$3\times3$ PPTES of rank four with the same range, then $\r=\r'$.
 \epp

\section{\label{sec:equivalence}
Equivalence classes of $3\times3$ PPTES of rank four}

The main objective of this section is to show that the set of
equivalence classes of $3\times3$ PPTES of rank four, equipped
with the quotient topology, is homeomorphic to the quotient
$R/A_5$ where $R\subset\bR^4$ is a product of four open
intervals and $A_5$ is the alternating group of order 60 acting
on $R$ by rational transformations. We also formulate a test for
checking the equivalence of two $3\times3$ PPTES of rank four.

Recall that $\cE'$ is the set of $3\times3$ PPTES of rank four,
and that $\cE'$ is $G$-invariant. The quotient space $\cE'/G$ parametrizes the set
of equivalence classes of $3\times3$ PPTES of rank four.
We equip $\cE'/G$ with the quotient topology and let
$\pi:\cE'\to\cE'/G$ be the projection map.

Let $\Omega=\{(a,b,c,d):a,b,c,d>0\}$ be the positive orthant
of $\bR^4$. Define the map $\omega:\Omega\to\cE'$ by the formula
$\omega(a,b,c,d)=C^\dag C$, where  $C=[C_0~C_1~C_2]$ and the
blocks $C_i$ are $4\times3$ matrices given by
Eq. (\ref{ea:Blokovi}). Our first objective is to prove that
the map $\pi\omega:\Omega\to\cE'/G$ is onto. This was proved in
our previous paper \cite{cd11JMP} apart from the fact that the
sign of the parameter $a$ was left ambiguous. The proof uses
the $G$-invariants of quintuples of product vectors which we now
briefly recall.

Let $(\ket{\a_i})_{i=0}^4$ be an (ordered) quintuple of vectors
in $\cH_A$ such that any three of them are linearly independent. To any such quintuple we assign three invariants
$(J_1^A,J_2^A,J_3^A)$.
These are certain complex numbers, different from 0 and 1,
subject to the relation $J_1^A J_2^A J_3^A=1$.
They are defined by the formulae
 \bea
J_1^A=\frac{\D_{2,0,4}\D_{0,1,3}}{\D_{2,0,3}\D_{0,1,4}}, \quad
J_2^A=\frac{\D_{0,1,4}\D_{1,2,3}}{\D_{0,1,3}\D_{1,2,4}}, \quad
J_3^A=\frac{\D_{1,2,4}\D_{2,0,3}}{\D_{1,2,3}\D_{2,0,4}},
 \eea
where $\D_{i,j,k}=\det[\ket{\a_i}~\ket{\a_j}~\ket{\a_k}]$. If two
quintuples, say $(\ket{\a_i})$ and $(\ket{\a'_i})$, have the same
invariants then there exists an invertible linear operator $V_A$
on $\cH_A$ such that $V_A\ket{\a_i}\propto\ket{\a'_i}$ for each
index $i$. (The proportionality constants may depend on the
index.) The converse is also valid.

To any quintuple of product vectors
$(\ket{\phi_i}=\ket{\a_i,\b_i})_{i=0}^4$ in $\cH$, which are in
general position, we assign six invariants
$(J_1^A,J_2^A,J_3^A,J_1^B,J_2^B,J_3^B)$, where the
$(J_1^A,J_2^A,J_3^A)$ are the invariants of the quintuple
$(\ket{\a_i})$ and $(J_1^B,J_2^B,J_3^B)$ are those of the quintuple
$(\ket{\b_i})$. If two quintuples of product vectors, say
$(\ket{\phi_i})$ and $(\ket{\psi_i})$, have the same invariants then
there exists an ILO, say $V$, such that
$V\ket{\phi_i}\propto\ket{\psi_i}$ for each index $i$, and the
converse holds.

For $\r\in\cE'$, $\ker\r$ contains exactly six product vectors,
say $\ket{\phi_i}$, $i=0,\ldots,5$. Moreover, these six product
vectors are in general position, see \cite[Theorem 22]{cd11JMP}.
There are in total 720 different sextuples
$(\ket{\phi_{\pi i}})$, where $\pi\in S_6$ is a permutation of the set $\{0,1,\ldots,5\}$.
We define the six invariants of such sextuple to be the
invariants of the quintuple which is obtained from the sextuple
by truncation, i.e., by omitting the last product vector
$\ket{\phi_{\pi 5}}$. It follows from \cite[Theorem 25]{cd11JMP}
that, for all sextuples $(\ket{\phi_{\pi i}})$, all six
invariants are real. Hence, each of
the invariants belongs to one of the open intervals $p=(0,1)$,
$P=(1,+\infty)$ and $N=(-\infty,0)$. To each of the sextuples we
associate a six letter {\em symbol}, with letters chosen from the
set $\{p,P,N\}$. The symbol is constructed from the sequence of
invariants $(J_1^A,J_2^A,J_3^A,J_1^B,J_2^B,J_3^B)$ by replacing
each number by the letter designating the open interval $p,P,N$
containing it. In the generic case, the 720 sextuples of product
vectors will have pairwise different sextuples of invariants.
However, only 12 different symbols arise. We refer to these
particular symbols as the {\em UPB symbols} (see \cite[Table
I]{cd11JMP}. Each of the 12 UPB symbols arises exactly 60 times.
For convenience we single out one of the 12 UPB symbols, namely
the symbol ppPNNp.

We assume (as we may) that the chosen sextuple $(\ket{\phi_i})$
is of {\em type} ppPNNp. The set of all permutations $\pi\in S_6$
for which $(\ket{\phi_{\pi i}})$ is of type ppPNNp is a subgroup
of $S_6$ isomorphic to the alternating group $A_5$ of order 60.
We shall refer to this subgroup as the {\em stabilizer} of the
symbol ppPNNp. One defines similarly
the stabilizers for other UPB symbols. All these stabilizers are
isomorphic to $A_5$ and are pairwise distinct. For the symbol
ppPNNp, the stabilizer is generated by the 5-cycle
$\a=(0,1,2,3,5)$ and the involution $\b=(1,2)(4,5)$.
Note that it permutes transitively the six product vectors
$\ket{\phi_i}$. From now on we
set $A_5=\langle \a,\b \rangle$.

Since $J_1^A J_2^A J_3^A=1$ and $J_1^B J_2^B J_3^B=1$, it
suffices to work with the following four invariants
$J_1^A,J_2^A,J_2^B,J_3^B$. For each $\pi\in A_5$, the sextuple
$(\ket{\phi_{\pi i}})$ has symbol ppPNNp and so the quadruple of
its invariants $(J_1^A,J_2^A,J_2^B,J_3^B)$ belongs to the open
infinite rectangular box $R=p\times p\times N\times p$ (a product
of four open intervals in the Euclidean space $\bR^4$). We shall
see that $R$ parametrizes the set of equivalence classes of
$3\times3$ PPTES $\r$ of rank four. However, this parametrization
is not one-to-one; generically it is sixty-to-one. The action of
$A_5$ on sextuples of product vectors in $\ker\r$ induces an
action of $A_5$ on $R$, and the equivalence classes are
represented by the orbits of $A_5$ in $R$. Let us describe
explicitly the action of $\a$ and $\b$ on the quadruple of
invariants $(a,b,c,d):=(J_1^A,J_2^A,J_2^B,J_3^B)$:
 \bea \label{Action5Cycle}
\a &:& (a,b,c,d)\to \left( \frac{(1-d)(b-c)}{(1-c)(1-abd)},~
b\cdot\frac{(1-c)(1-acd)}{(1-cd)(b-c)},~
-\frac{1}{d}\cdot\frac{(1-b)(1-acd)}{(1-a)(b-c)},~
\frac{(1-a)(1-abd)}{(1-ab)(1-acd)} \right); \\
\b &:& (a,b,c,d)\to \left( d\cdot\frac{1-c}{1-cd},~
\frac{(1-ab)-ce}{(1-c)(1-abd)},~
\frac{b}{c}\cdot\frac{(1-ab)-ce}{(1-b)(1-abd)},~
-c\cdot\frac{(1-a)(1-abd)}{(1-ab)-ce} \right),
 \eea
where $e=(1-a)+ad(1-b)$. The meaning of, say, the first formula
is that its right hand member is the quadruple of invariants of
the sextuple $(\ket{\phi_{\a i}})=(\ket{\phi_5},\ket{\phi_0},
\ket{\phi_1},\ket{\phi_2},\ket{\phi_4},\ket{\phi_3})$.

From the preceding discussion we obtain the following test for
equivalence.
 \bt \label{thm:EqTest}
Let $\r,\r'\in\cE'$ and let $(\ket{\phi_i})$ and $(\ket{\phi'_i})$
be sextuples of product vectors in $\ker\r$ and $\ker\r'$,
respectively. Then $\r$ and $\r'$ are SLOCC equivalent if and only if
there exists a permutation $\pi\in S_6$ such that the sextuples
$(\ket{\phi_{\pi i}})$ and
$(\ket{\phi'_i})$ have the same invariants.
 \et
 \bpf
{\em Necessity.} Let $\r'=V\r V^\dag$, where $V$ is an ILO. Since
$\r\ket{\phi_i}=0$ and $\ker\r=V^\dag\ker\r'$, we have
$V^\dag\ket{\phi'_i}=c_i\ket{\phi_{\pi i}}$ for some permutation
$\pi$ and some nonzero scalars $c_i$. It follows that the sextuples
$(\ket{\phi_{\pi i}})$ and $(\ket{\phi'_i})$ have the same
invariants.

{\em Sufficiency.} Since $(\ket{\phi_{\pi i}})$ and
$(\ket{\phi'_i})$ have the same invariants, there exists an ILO, say
$V$, such that $V^\dag\ket{\phi'_i}=c_i\ket{\phi_{\pi i}}$ for some
permutation $\pi$ and some nonzero scalars $c_i$. Since the
$\ket{\phi'_i}$ span $\ker\r'$ and the $\ket{\phi_i}$ span $\ker\r$,
we have $V^\dag\ker\r'=\ker\r=V^\dag\ker(V\r V^\dag)$. It follows
that the states $\r$ and $V\r V^\dag$ have the same kernel and
range. By Proposition \ref{prop:3x3rank4PPTES,samerange} these
two states are equivalent, and so are $\r$ and $\r'$.
 \epf

We shall now prove that each point of $R$ is the quadruple of
invariants of some $\r\in\cE'$ and that we can indeed require in
Proposition \ref{prop:3x3rank4PPTES,invariantexpression} that
$a>0$. In view of Theorem \ref{thm:EqTest} this implies that the
map $\pi\omega:\Omega\to\cE'/G$ is onto.

 \bt \label{thm:ParPPTES}
Every $\s\in\cE'$ is SLOCC equivalent to a state 
$\r=\omega(a,b,c,d)$ (where all $a,b,c,d>0$).
For each point $(x,y,z,w)\in R$ there exists a state $\r\in\cE'$
such that $(x,y,z,w)=(J_1^A,J_2^A,J_2^B,J_3^B)$ for some sextuple
of product vectors in $\ker\r$.
 \et
 \bpf
To prove the first assertion we choose a sextuple of product
vectors $(\ket{\phi_i})_{i=0}^5$ in $\ker\s$ with invariants
of type $ppPNNp$. Thus we have
$(J_1^A,J_2^A,J_2^B,J_3^B)\in R$. Let us define the positive numbers
$b,c,d$ by
 \bea \label{iz-b2}
b^2 &=& -\frac{J_2^BJ_3^B}{J_2^A}\cdot \frac{(1-J_2^A)(1-J_1^A
J_2^A)}
{(1-J_2^B J_3^B)(1-J_1^A J_2^B J_3^B)}, \\
c^2 &=& -\frac{1}{J_2^B}\cdot \frac{(1-J_1^A J_2^B)(J_2^A-J_2^B
J_3^B)}
{(1-J_3^B)(1-J_1^A J_2^A)}, \\
d^2 &=& J_1^A\cdot \frac{(1-J_2^B)(1-J_3^B)} {(1-J_1^A)(1-J_1^A
J_2^B J_3^B)}.
 \eea
(It is easy to see that the right hand sides are positive.)

We can now define $a>0$ by the formula
 \bea \label{iz-a}
a &=& \frac{bcd}{J_3^B}\cdot \frac{(1-J_1^A)(1-J_1^A J_2^B
J_3^B)(J_2^A-J_2^B J_3^B)} {(1-J_2^A)(1-J_1^A J_2^B)^2}.
 \eea

We claim that $\r:=\omega(a,b,c,d)$ and $\s$ are equivalent.
It suffices to show that we can choose a sextuple of product
vectors in $\ker\r$ having the same invariants as the sextuple $(\ket{\phi_i})$. We observe that $\ker\r$ contains the product
vectors $\ket{ii}$ for $i=0,1,2$. We have to find the remaining
three product vectors in this kernel. We proceed as in the proof
of \cite[Theorem 25]{cd11JMP}.
We introduce the cubic polynomial
 \bea
f(z) &=& abz(cz-1-d^2)(c-(1+c^2)z)+d(cz-1)(b^2c-(1+b^2+b^2c^2)z).
 \eea
It has three real roots $z_1,z_2,z_3$ such that
 \bea
z_3<0,\quad \l<z_1<c/(1+c^2), \quad 1/c<z_2<(1+d^2)/c,
 \eea
where $\l=b^2c/(1+b^2+b^2c^2)$. Explicitly, they are given by the
formulae
 \bea
z_1 &=& \frac{J_3^B}{c}\cdot
\frac{1-J_1^A J_2^B}{(1-J_1^A J_2^B J_3^B)}, \\
z_2 &=& \frac{1}{c}\cdot
\frac{1-J_1^A J_2^B}{(1-J_1^A J_2^B J_3^B)}, \\
z_3 &=& -\frac{1}{c}\cdot \frac{(1-J_2^A)(1-J_1^A
J_2^B)}{(1-J_1^A)(J_2^A-J_2^B)}.
 \eea
[The verification that these are indeed the roots of $f(z)$ is
a very tedious job if done by hand, but it is trivial for a
package for symbolic algebraic computations such as Maple.]

The three additional product vectors in the kernel of $\r$ are
$\ket{\psi_i}$ $(i=1,2,3)$ defined by \cite[Eq. (110)]{cd11JMP}.
By using the above expressions for $a,b,c,d$ and $z_1,z_2,z_3$
and the formulae \cite[Eqs. (119-120)]{cd11JMP}, we can compute
the invariants of the sextuple $(\ket{00},\ket{11},\ket{22},
\ket{\psi_1},\ket{\psi_2},\ket{\psi_3})$.
We obtain that they are the same as the invariants of the
sextuple $(\ket{\phi_i})$. This proves our claim and completes
the proof of the first assertion.

To prove the second assertion, let us define a smooth map
$\Phi:R\to\Omega$. We set $\Phi(x,y,z,w)=(a,b,c,d)$, where $a,b,c,d$
are defined by the formulae (\ref{iz-b2})-(\ref{iz-a}) in which we
replace $J_1^A,J_2^A,J_2^B,J_3^B$ with $x,y,z,w$, respectively. The
proof of the first assertion shows that the quadruple of invariants
of the state $\r$, constructed above, is exactly $(x,y,z,w)$. Hence,
the second assertion is proved.
 \epf

It is not hard to show that the above map $\Phi$ is a local
diffeomorphism. We believe that it is a (global) diffeomorphism,
but this question is beyond the scope of this paper.
However, let us observe that if
$(x,y,z,w)\in\Phi^{-1}(a,b,c,d)$, then $(x,y,z,w)$ is a
quadruple of invariants of some ordered sextuple of the
six product vectors in the kernel of $\omega(a,b,c,d)$.
Moreover, the type of this sextuple is ppPNNp.
Consequently, the fibre $\Phi^{-1}(a,b,c,d)$ is finite of
cardinality at most 60.

 \bcr
The set, $\cE$, of all normalized $3\times3$ PPTES of rank
four on $\cH$ has dimension 36.
 \ecr
 \bpf
This set is a real semialgebraic set and so it has a well defined
dimension, see e.g. \cite{Boch98}. For convenience we shall work
with non-normalized states, and so the dimension will increase
by one. More precisely, we have $\cE'=\{\l\r:\l>0,\r\in\cE\}$.
Denote by $X$ the image of $\omega$,
$X=\{\omega(a,b,c,d):(a,b,c,d)\in\Omega\}\subseteq\cE'$.
Each equivalence class $E\subseteq\cE'$ is an orbit of $G$. It follows from \cite[Proposition 27]{cd11JMP} that the identity
component of the stabilizer of $\r\in X$ is the 3-dimensional
subgroup $\{(z_1I_3,z_2I_3):|z_1z_2|=1\}$ of $G$.
Hence, $\dim E=\dim G-3=33$. Each equivalence class $E$ intersects $X$ in at least one and at most 60 points.
Consequently, $\dim\cE'=\dim X+\dim E=37$. The normalization
decreases the dimension by 1, thus $\dim\cE=36$.
 \epf

We point out that this agrees with the computation performed in
\cite{hhm11}.

Let $p:R\to R/A_5$ be the canonical projection map onto the quotient
of $R$ by the action of $A_5$. As usual, the quotient space $R/A_5$
is equipped with the quotient topology. We cannot define a map
$\cE'\to R$ because there is no natural choice of ordering the six
product vectors in the kernel of a state $\r\in\cE'$. However, this
ambiguity disappears when we replace $R$ by $R/A_5$ and specify that
the sextuple of product vectors in $\ker\r$ is chosen to be of type
ppPNNp. Thus we obtain a continuous map $q:\cE'\to R/A_5$. We now
use the universal property of the quotient maps $\pi:\cE'\to\cE'/G$
and $p:R\to R/A_5$ to prove the following.

 \bt \label{Klasif}
The quotient spaces $\cE'/G$ and $R/A_5$ are homeomorphic.
 \et
 \bpf
Since the map $q:\cE'\to R/A_5$ is continuous and is constant
on the fibers of the map $\pi:\cE'\to\cE'/G$, there exists a unique
continuous map $f:\cE'/G\to R/A_5$ such that $q=f\pi$.
At the end of the proof of the previous theorem, we have
introduced the smooth map $\Phi:R\to\Omega$.
Since the map $\pi\omega\Phi:R\to\cE'/G$ is continuous and
constant on the fibers of the map $p:R\to R/A_5$, there
exists a unique continuos map $g:R/A_5\to\cE'/G$ such that
$\pi\omega\Phi=gp$. The universal property also implies that
$fg$ and $gf$ are the identity maps. Hence, $f$ is a
homeomorphism.
 \epf

We shall now examine the two classical examples of UPB,
namely {\bf Pyramid} and {\bf Tiles}.

 \bex \label{Piramida} {\rm
The quadruple of invariants for the {\bf Pyramid} UPB is
$(-(\sqrt{5}-1)/2,(3+\sqrt{5})/2,(3-\sqrt{5})/2,(\sqrt{5}+1)/2)$
with symbol NPNPpP. If we rearrange the five vectors of this UPB in
order $[2,1,4,3,5]$, the quadruple of invariants becomes
$((\sqrt{5}-1)/2,(\sqrt{5}-1)/2,-(\sqrt{5}+1)/2,(3-\sqrt{5})/2)$. So
it has the desired symbol ppPNNp and the point with these
coordinates is in the region $R$.

We claim that this point is the unique fixed point of the
transformation $\a$ as given by Eq. (\ref{Action5Cycle}).
Indeed, by equating the two members of Eq. (\ref{Action5Cycle}),
we obtain a system of four equations. The first two equations
can be solved easily and give
 \bea
c=\frac{1-b-ab+a^2b^2}{(1-a)(1-a-b)}, \quad
d=\frac{a+b-1}{1-b(1-ab)}.
 \eea
Then the remaining two equations factorize, and after droping
the factors that cannot vanish when $a,b\in(0,1)$, we obtain
the system of two equations:
 \bea
&& a^2b^3-ab^2+ab-a-b+1=0, \\
&& a^2b^3-a^2b^2-a^2b-2ab^2+3ab+b-1=0.
 \eea
The resultant of these two polynomials with respect to the
variable $a$ is $b(1+b^2)(1-b)^3(1-b-b^2)$. It has a unique root
in the interval $(0,1)$ namely $b=(\sqrt{5}-1)/2$. The proof of
our claim can now be completed easily.

Moreover, one can verify that the transformation $\b$ fixes the
unique fixed point of $\a$. Hence, this point is fixed by all
elements of $A_5$. Note that this observation agrees with the
fact mentioned in \cite{cd11JMP} that the stabilizer of
{\bf Pyramid} PPTES in the group ${\rm PGL_3}\times{\rm PGL_3}$
is the alternating group $A_5$.
\hfill $\square$ }
 \eex

 \bex \label{Ploca} {\rm
Let us now consider the {\bf Tiles} UPB. The quadruple of its
invariants is $(-1/2,2,1/3,3/2)$ again with symbol NPNPpP.
By rearranging the five vectors in order $[2,1,4,3,5]$,
we obtain the quadruple of invariants $(1/2,2/3,-1,1/2)$
having the desired symbol ppPNNp. The orbit of $A_5$ containing
the point $(1/2,2/3,-1,1/2)\in R$ has size five.
The other four points of this orbit are $(1/2,1/2,-2,1/4)$,
$(2/3,1/2,-2,1/2)$, $(2/3,3/4,-3,1/3)$ and $(3/4,2/3,-1,1/3)$.
Each of these five points $(x,y,z,w)$ gives the parameters
$(a,b,c,d)=\Phi(x,y,z,w)\in\Omega$ such that the state
$\omega(a,b,c,d)\in\cE'$ is equivalent to the PPTES obtained
from {\bf Tiles}.
For instance, by substituting the coordinates of the first point
for the invariants $J_1^A,J_2^A,J_2^B,J_3^B$ in
Eqs. (\ref{iz-b2}-\ref{iz-a}), we obtain that $a=7\sqrt{21}/27$,
$b=2/3\sqrt{5}$, $c=\sqrt{21}/2$ and $d=2\sqrt{5}$.
\hfill $\square$ }
 \eex

At the referee's suggestion, we shall also consider the
one-parameter generalization of Choi's PPTES.

 \bex \label{Choi} {\rm
This one-parameter family $\r_\l$ of $3\times3$ PPTES of rank four is given in \cite[p. 169]{hkp03} by its density matrix
\bea
A=\left[ \begin{array}{ccccccccc}
1 &      &         &         & 1 &      &      &         & 1 \\
  & \l^2 &         &         &   &      &      &         &   \\
  &      & \l^{-2} &         &   &      &  1   &         &   \\
  &  1   &         & \l^{-2} &   &      &      &         &   \\
1 &      &         &         & 1 &      &      &         & 1 \\
  &      &         &         &   & \l^2 &      &    1    &   \\
  &      &   1     &         &   &      & \l^2 &         &   \\
  &      &         &         &   &  1   &      & \l^{-2} &   \\
1 &      &         &         & 1 &      &      &         & 1
\end{array} \right].
\eea
(The blank entries are zeros.) The parameter $\l$ is assumed to
be positive and different from 1. For simplicity, we shall assume that $0<\l<1$. The six product vectors in $\ker\r_\l$
are the tensor products of the corresponding columns in the
following two $3\times6$ matrices
\bea
\left[ \begin{array}{rrrrrr}
1 & 0 & \l & 1 & 0 & -\l \\
\l & 1 & 0 & -\l & 1 & 0 \\
0 & \l & 1 & 0 & -\l & 1 \end{array} \right], \quad
\left[ \begin{array}{rrrrrr}
-\l & 0 & 1 & \l & 0 & 1 \\
1 & -\l & 0 & 1 & \l & 0 \\
0 & 1 & -\l & 0 & 1 & \l \end{array} \right].
\eea
This sextuple of product vectors has symbol pNNPpp. After
interchanging the third and fourth columns of both matrices,
the symbol becomes ppPNNp and a computation shows that the
quadruple of invariants of the reordered sextuple is given by
\bea
J_1^A=\frac{1+\l^3}{2},\quad
J_2^A=\frac{1-\l^3}{1+\l^3},\quad
J_2^B=-\frac{1+\l^3}{1-\l^3},\quad
J_3^B=\frac{2\l^3}{1+\l^3}.
\eea
By applying the formulae \eqref{iz-b2}-\eqref{iz-a}, we conclude that $\r_\l$ is  equivalent to the state $\s$ defined in
Proposition \ref{prop:3x3rank4PPTES,invariantexpression}, with
$a,b,c,d>0$ given by
\bea
b^2=\frac{2\l^6}{1+\l^6},\quad
c^2=\frac{(3+\l^6)(1+3\l^6)}{(1-\l^6)^2},\quad
d^2=\frac{2}{1+\l^6},\quad
a=\frac{bcd}{2\l^6}\cdot\frac{(1-\l^{12})(1+3\l^6)}{(3+\l^6)^2}.
\eea
\hfill $\square$ }
 \eex

Finally we show a simple result on PPTES.

 \bl
Given $\r\in\cE'$, there is no $3\times3$ extreme state $\s$ of
rank five, such that $\cR(\s) = \ker\rho$.
 \el
 \bpf
Suppose such $\s$ exists. We may assume that $\r=\omega(a,b,c,d)$
for some $a,b,c,d>0$. Using Eq. \eqref{ea:Blokovi}, the product
vector $\ket{00}$ belongs to $\cR(\s)$ and $\cR(\s^\G)$. So $\s$ is
not an edge state, and we have reached a contradiction.
 \epf

\section{\label{sec:checkerboard}
Checkerboard states}

One of the early examples of PPTES was the multi-parameter family of
checkerboard states constructed in \cite{BrPeres}, see also
\cite{djokovic11}. We will show that, up to equivalence, all
checkerboard PPTES can be parametrized by just two real parameters.
By using invariants, we have devised a test for checking whether an
arbitrary $3\times3$ PPTES of rank four is equivalent to a
checkerboard state.

We define the checkerboard states (see \cite[section 7]{cd11JPA}) to
be the states $\r$ which can be written as $\r=C^\dg C$, where
$C=[C_1~C_2~C_3]$ and the $C_i$ are complex matrices having the
following form:
 \bea \label{CheckerBoard}
C_1=\left(\begin{array}{ccc}
 a&0&d\\0&g&0\\j&0&m\\0&q&0\end{array}\right),\quad
C_2=\left(\begin{array}{ccc}
 0&c&0\\f&0&i\\0&l&0\\p&0&s\end{array}\right),\quad
C_3=\left(\begin{array}{ccc}
 b&0&e\\0&h&0\\k&0&n\\0&r&0\end{array}\right).
 \eea

 \bpp \label{KanCheckerBoard}
Every checkerboard PPTES is equivalent to one with
 \bea \label{CheckerBoardSpecial}
C_1=\left(\begin{array}{ccc}
 1&0&0\\0&u&0\\0&0&0\\0&0&0\end{array}\right),\quad
C_2=\left(\begin{array}{ccc}
 0&u&0\\1&0&0\\0&1&0\\0&0&1\end{array}\right),\quad
C_3=\left(\begin{array}{ccc}
 0&0&0\\0&v&0\\v&0&1\\0&1&0\end{array}\right),\quad u,v>0.
 \eea
Consequently, there exist $3\times3$ PPTES of rank four not
equivalent to any checkerboard state.
 \epp
 \bpf
Let $\r=C^\dag C$ be a $3\times3$ PPTES of rank four, where
$C=[C_1~C_2~C_3]$ with the $C_i$ given by (\ref{CheckerBoard}). Then
any linear combination of the $C_i$ must have rank at least two. By
replacing $C_1$ with a suitable linear combination of $C_1$ and
$C_3$, we can assume that $am=dj$. By applying an ILO and
premultiplying $C$ by a unitary matrix, we may assume that
 \bea
C_1=\left(\begin{array}{ccc}
 1&0&0\\0&1&0\\0&0&0\\0&0&0\end{array}\right),\quad
C_2=\left(\begin{array}{ccc}
 0&c&0\\f&0&i\\0&l&0\\p&0&s\end{array}\right),\quad
C_3=\left(\begin{array}{ccc}
 b&0&e\\0&h&0\\k&0&n\\0&r&0\end{array}\right).
 \eea
Since $\r^\G\ge0$, we must have $e=i=0$. By using another ILO, we
can further simplify these matrices to obtain
 \bea
 C_1=\left(\begin{array}{ccc}
 1&0&0\\0&1&0\\0&0&0\\0&0&0\end{array}\right),\quad
 C_2=\left(\begin{array}{ccc}
 0&c&0\\1&0&0\\0&l&0\\0&0&1\end{array}\right),\quad
 C_3=\left(\begin{array}{ccc}
 0&0&0\\0&h&0\\k&0&1\\0&r&0\end{array}\right).
 \eea
As $\r$ is entangled, $\cR(\r)$ must be a CES and so $chlr\ne0$. By
using an argument from the proof of \cite[Theorem 28]{cd11JPA}, we
obtain that $h=rk^*$, $|c|=1$ and $l=cr^*k/k^*$. Thus
 \bea
 C_1=\left(\begin{array}{ccc}
 1&0&0\\0&1&0\\0&0&0\\0&0&0\end{array}\right),\quad
 C_2=\left(\begin{array}{ccc}
 0&c&0\\1&0&0\\0&cr^*k/k^*&0\\0&0&1\end{array}\right),\quad
 C_3=\left(\begin{array}{ccc}
 0&0&0\\0&rk^*&0\\k&0&1\\0&r&0\end{array}\right).
 \eea
Let us choose $u_2$ such that $u_2^2=c$ and premultiply $C$ with the
unitary matrix $\diag(1,u_2,u_3,u_4)$, where $u_3=|r|k^*/r^*k$ and
$u_4=u_2|k|/k$. Next we postmultiply each $C_i$ with
$\diag(1,u_2^*/|r|,|k|/k^*)$, and multiply $C_2$ and $C_3$ with
$u_2^*$ and $|rk|/rk^*$, respectively. (Note that these
transformations form an ILO.) Now we have
 \bea
 C_1=\left(\begin{array}{ccc}
 1&0&0\\0&1/|r|&0\\0&0&0\\0&0&0\end{array}\right),\quad
 C_2=\left(\begin{array}{ccc}
 0&1/|r|&0\\1&0&0\\0&1&0\\0&0&1\end{array}\right),\quad
 C_3=\left(\begin{array}{ccc}
 0&0&0\\0&|k|&0\\|k|&0&1\\0&1&0\end{array}\right).
 \eea
Thus the first assertion is proved, and the second follows from the
first because the matrices (\ref{CheckerBoardSpecial}) depend only
on two positive parameters.
 \epf

We can now characterize the checkerboard PPTES up to equivalence.
 \bpp \label{pp:karakterizacija}
A $3\times3$ PPTES of rank four is equivalent to a checkerboard
state if and only if, for some ordering of the product vectors in
its kernel, its invariants are of type PNNpPP and satisfy the
equations $J_2^A=J_3^A$ and $J_2^B=J_3^B$.
 \epp
 \bpf
{\em Necessity.} Let $\r$ be a $3\times3$ checkerboard PPTES of rank
four and $(J_1^A,J_2^A,J_3^A,J_1^B,J_2^B,J_3^B)$ its invariants with
symbol PNNpPP. By Proposition \ref{KanCheckerBoard} we have
$\r=C^\dag C$ where $C=[C_1~C_2~C_3]$ and the $C_i$ are given by
(\ref{CheckerBoardSpecial}). We have to find the six product vectors
in $\ker\r$. First observe that the product vectors
$\ket{\phi_0}=\ket{02}$ and $\ket{\phi_5}=\ket{20}-v\ket{22}$ belong
to $\ker\r$. Let us exhibit the remaining four.

Let $\pm x_1,\pm x_2$ be the roots of the biquadratic polynomial
$f(x)=u^2(1+v^2)x^4-(u^2v^2+2u^2+v^2)x^2+u^2$. Since $f(1)=-v^2<0$,
we may assume that $0<x_1<1<x_2$. For any real $t$ let
$\ket{\phi(t)}=\ket{\a(t)}\ox\ket{\b(t)}$, where
$\ket{\a(t)}=v\ket{0}+tv\ket{1}+u(t^2-1)\ket{2}$ and
$\ket{\b(t)}=uvt^2\ket{0}-tv\ket{1}+(t^2-1)\ket{2}$. One can easily
verify that the product vectors $\ket{\phi_1}=\ket{\phi(x_1)}$,
$\ket{\phi_2}=\ket{\phi(-x_1)}$, $\ket{\phi_3}=\ket{\phi(x_2)}$,
$\ket{\phi_4}=\ket{\phi(-x_2)}$ belong to $\ker\r$.

A computation shows that the invariants of the sextuple
$(\ket{\phi_i})_{i=0}^5$ are $(1/\m^2,-\m,-\m,1/\l^2,\l,\l)$, where
$\l=(x_2+x_1)/(x_2-x_1)>1$ and $\m=\l(1-x_1x_2)/(1+x_1x_2)<\l$. As
$x_1^2 x_2^2=1/(1+v^2)<1$, we have $\mu>0$. {}From the equality
$x_1^2=(\l-\m)(\l-1)/(\l+\m)(\l+1)$ we deduce that $\m<1$. Hence,
the invariants indeed have the required symbol PNNpPP and the
equalities $J_2^A=J_3^A$ and $J_2^B=J_3^B$ hold.

{\em Sufficiency.} Now assume that $\s$ is a $3\times3$ PPTES of
rank four and that $(J_1^A,J_2^A,J_3^A,J_1^B,J_2^B,J_3^B)$ are the
invariants, having the symbol PNNpPP, of some sextuple of product
vectors in $\ker\s$. We also assume that $J_2^A=J_3^A$ and
$J_2^B=J_3^B$, and so we can write these invariants as
$(1/\m^2,-\m,-\m,1/\l^2,\l,\l)$, where $\l>1$ and $\m>0$. Since the
first letter of the above symbol is P, we have $\m<1$. We set
 \bea
v=\frac{2\sqrt{\l\m}}{\l-\m}>0, \quad
w=2\cdot\frac{\l-\m}{\l+\m}\cdot\frac{\l^2+1}{\l^2-1}.
 \eea
It is not hard to verify that
 \bea
(1+v^2)(w-1)-1=\frac{4\l^2(1-\m^2)}{(\l^2-1)(\l-\m)^2}>0,
 \eea
and so we can define $u>0$ by the equation
$u^2((1+v^2)(w-1)-1)=v^2$.

Let $\r=C^\dag C$ be the checkerboard state where $C=[C_1~C_2~C_3]$
and the $C_i$ are given by (\ref{CheckerBoardSpecial}) with the
parameters $u$ and $v$ as defined above. The biquadratic polynomial
$f(x)=u^2(1+v^2)x^4-(u^2v^2+2u^2+v^2)x^2+u^2$ has roots $\pm x_1,\pm
x_2$, where $x_1,x_2>0$ are given by
 \bea
x_1^2=\frac{(\l-\m)(\l-1)}{(\l+\m)(\l+1)}, \quad
x_2^2=\frac{(\l-\m)(\l+1)}{(\l+\m)(\l-1)}.
 \eea
The computation performed in the first part of the proof shows that
$(1/\m^2,-\m,-\m,1/\l^2,\l,\l)$ are also the invariants of a
sextuple of product vectors in $\ker\r$. Hence, $\r$ and $\s$ are
equivalent.
 \epf

We remark that if we use the invariants of type ppPNNp (instead of
PNNpPP) then the two equations in the proposition should be replaced
by $J_2^A+J_3^A=2J_2^A J_3^A$ and $J_2^B+J_3^B=2J_2^B J_3^B$. Since
in Theorems \ref{thm:EqTest},\ref{thm:ParPPTES} we use only the
invariants $J_1^A,J_2^A,J_2^B$ and $J_3^B$, we mention that the
former equation can be rewritten as $J_2^A(1-J_1^AJ_2^A)=1-J_2^A$.

\section{\label{sec:extreme} Extreme and edge $3\times 3$ states}

In recent years extreme and edge $3\times 3$ states have been
extensively studied. A primary reason is that PPTES do not exist in
any bipartite space of smaller dimension \cite{hhh96,hst99,hlv00}.
On the other hand, many important problems are open for $3\ox3$
systems. For example, a $3\times3$ PPT state of rank four is
separable if and only if there is a product vector in the range of
this state \cite{cd11JPA}. But there is no analytical criterion for
deciding which $3\times3$ states of rank larger than four are
separable. It is also unknown whether two-qutrit entangled states of
rank larger than three are distillable \cite{cc08,cd11JPA}.
Characterizing $3\times 3$ extreme and edge PPTES may provide a
better understanding of these problems.

We begin with extreme states. There is a simple necessary
condition for extremality of states \cite{lmo07}.
 \bpp \label{prop:extremeNecCond}
Let $\r$ be an $M\times N$ PPT state of birank $(r,s)$. If
$r^2+s^2>M^2 N^2+1$ then $\r$ is not extreme.
 \epp

By this proposition, the birank of an extreme $3\times3$ PPTES $\r$ must be (up to ordering) one of the following five pairs:  \bea
 \label{ea:5typeEXTREME}
 (4,4),~ (5,5),~ (6,6),~ (5,6),~ (5,7).
 \eea
Examples for the first three pairs have been known for some time. First, the extreme states $\r$ of birank $(4,4)$ were thoroughly analyzed and characterized by UPB \cite{cd11JMP,skowronek11}.
Second, the edge states $\r$ of biranks $(5,5)$ and $(6,6)$ have
been analytically constructed \cite{clarisse06}. They were proved
to be extreme in Ref. \cite{kk07,ha09}. Third, families of edge
states $\r$ for the remaining two pairs, $(5,6)$ and $(5,7)$,
have been constructed very recently \cite{ko12}.
In Table {\bf I} at the end of this section, we will show that for suitable parameter values, these edge states turn out to
be extreme. They are the first examples of extreme states $\r$ with $\rank\r\ne\rank\r^\G$. Hence, each of the pairs
\eqref{ea:5typeEXTREME} is the birank of some $3\times3$ extreme state.

Next we consider the $3\times 3$ edge states $\r$ of birank $(r,s)$.

By \cite{cd11JMP} we have $r=4$ if and only if $s=4$. It is known
that $r+s\le14$ \cite{kkl11,ko12}.
(It has been claimed in \cite{sbl01} that $r+s\le13$.
However, it was shown in \cite{kkl11,ko12} that this
claim is false.)
The birank $(r,s)$ cannot
be equal to $(5,9)$ because any 5-dimensional subspace of $\cH$
contains at least one product vector.

We conclude that, up to ordering, $(r,s)$ must be one of the
following eight pairs:
 \bea
 \label{ea:8typeEDGE}
 (4,4),~ (5,5),~ (6,6),~ (5,6),~ (5,7),~ (5,8),~ (6,7),~ (6,8).
 \eea
Recall that there exist extreme states for the first five pairs.
As any extreme state of rank bigger than one is also an edge
state, it remains to consider only the last three pairs in
this list. Such edge states have been constructed recently
in \cite{ko12}.
Since the examples for the first five pairs  in \eqref{ea:8typeEDGE} are extreme states, it is an interesting
question to ask whether in these five cases there exist
$3\times3$ edge states $\r$ which are not extreme.

We have constructed such states $\r$ (see Table {\bf I}) for
pairs $(6,6)$, $(5,6)$ and $(5,7)$.
Since every PPTES of birank $(4,4)$ is extreme, only the case $(5,5)$ remains in doubt. We have tested a few known $3\times3$
edge states of birank $(5,5)$, but all of them turned out to be extreme (see Table {\bf I}).
We conjecture that any $3\times3$ edge state of birank $(5,5)$
is extreme.
This should be compared with the known result that any
$2\times4$ PPTES of birank $(5,5)$ is extreme \cite{agk10}.
The $3\times3$ case is more challenging because there exist
$3\times3$ PPTES of birank $(5,5)$ which are not edge states.
An example is $\r=\s+\e\proj{00}$ where $\s$ is a $3\times3$
PPTES of birank $(4,4)$, and $\e>0$ is sufficiently small.

Another interesting problem is whether our results can be
extended to higher dimensions. For instance, examples of extreme $M\times N$ PPTES of rank $M+N-2$ have been constructed
recently \cite{cd12}.
One may ask how many parameters are necessary to describe
such states.

 \label{item:extremetable}
 Table {\bf I} of $3\times3$ edge and extreme states $\r$.  \\(Computations performed using Maple.)
 \bem

\item
 Birank $(5,5)$
 \bem
\item
 \cite[Section II]{clarisse06}
 \\Extreme state.
 \\$\ker\r$: 2 product vectors.
 \\$\cR(\r)$: Infinitely many product vectors.
 \\$\cR(\r^\G)$: Infinitely many product vectors.
\item
 \cite{hk05,kk07} with  $s=1$ and $t=2$.
 \\Extreme state.
 \\$\ker\r$: 2 product vectors.
 \\$\cR(\r)$: Infinitely many product vectors.
 \\$\cR(\r^\G)$: Infinitely many product vectors.
\item
 \cite[Sec. 4]{ko12} with $b=2$ and $\theta=\pi/6$.
 \\Extreme state.
 \\$\ker\r$: CES.
 \\$\cR(\r)$: 6 product vectors.
 \\$\cR(\r^\G)$: 6 product vectors.
 \eem

\item
 Birank $(6,5)$
 \bem
\item
 \cite{hk05} with $s=2$,
 \\Edge state, but not extreme,
 \\$\ker\r$: 2 product vectors.
 \\$\cR(\r^\G)$: 3 product vectors.
\item
 \cite[Sec. 4]{ko12} with $b=2$, $\theta=\pi/4$ and $r=0$.
 \\Extreme state.
 \\$\ker\r$: CES.
 \\$\cR(\r^\G)$: 6 product vectors.
 \eem

\item
 Birank $(6,6)$
 \bem
\item
 \cite[Section III]{clarisse06}
 \\Extreme state.
 \\$\ker\r$: CES.
\item
 \cite[Sec. 4]{ko12} with $b=2$, $\theta=\pi/6$, $(\xi|\eta)=0$
and $(\eta|\zeta)=(\zeta|\xi)=1$.
 \\Edge state, but not extreme.
 \\$\ker\r$: CES.
 \eem

\item
 Birank $(7,5)$
 \bem
\item
 \cite{hk05} with $s=2$.
 \\Edge state, but not extreme.
 \\$\ker\r$: CES.
 \\$\cR(\r^\G)$: 6 product vectors.
\item
 \cite[Sec. 4]{ko12} $r=3/7$, $b=2$ and $\theta=\arccos(9/14)$.
 \\Extreme state.
 \\$\ker\r$: CES.
 \\$\cR(\r^\G)$: 6 product vectors.
 \eem

\item
 Birank $(7,6)$
 \bem
\item
 \cite{hk05} with $s=2$.
 \\Edge state, but not extreme.
 \\$\ker\r$: CES.
 \eem

\item
 Birank $(8,5)$
 \bem
\item
 \cite[Sec. 4]{ko12} with $b=2$, $\theta=\pi/4$ and $r=3/7$.
 \\Edge state, but not extreme.
 \\$\ker\r$: CES.
 \\$\cR(\r^\G)$: 6 product vectors.
 \eem

\item
 Birank $(8,6)$
 \bem
\item
 \cite[Sec. 3]{ko12} with $b=2$ and $\theta=\pi/4$.
 \\Edge state, but not extreme.
 \\$\ker\r$: CES.
\eem

 \eem

\acknowledgments

We thank the anonymous referee for his comments and suggestions. The
first author was mainly supported by MITACS and NSERC. The CQT is
funded by the Singapore MoE and the NRF as part of the Research
Centres of Excellence programme. The second author was supported in
part by an NSERC Discovery Grant.

%\acknowledgments
%
%The first author was mainly supported by
%MITACS and NSERC. The CQT is funded by the Singapore MoE and the NRF
%as part of the Research Centres of Excellence programme. The second
%author was supported in part by an NSERC Discovery Grant.


\begin{thebibliography}{99}

\bibitem{agk10} R. Augusiak, J. Grabowski, M. Kus, and M.
Lewenstein, Searching for extremal PPT entangled states, Optics
Commun. {\bf 283} (2010), 805-813.

\bibitem{bbc93} C.H. Bennett, G. Brassard, C. Crepeau, R. Jozsa, A. Peres, and W. Wootters,
Teleporting an unknown quantum state via dual classical and
einstein-podolsky-rosen channels, Phys. Rev. Lett. {\bf 70}, 1895
(1993).

\bibitem{bdm99} C.H. Bennett, D.P. DiVincenzo, T. Mor, P.W. Shor,
J.A. Smolin, and B.M. Terhal, Unextendible product bases and bound
entanglement, \prl {\bf82} (1999), 5385-5388.

\bibitem{bell64} J. Bell, On the Einstein-Podolsky-Rosen paradox,
Physics {\bf1}, 195 (1964).

\bibitem{Boch98} J. Bochnak, M. Coste, and M.-F. Roy, Real Algebraic
Geometry, New York, Springer, 1998.


\bibitem{BrPeres} D. Bru\ss, and A. Peres,
Construction of quantum states with bound entanglement, \pra {\bf
61}, 030301(R) (2000).

\bibitem{cc08} Lin Chen and Yi-Xin Chen, Rank-three bipartite entangled states are distillable, \pra {\bf 78}, 022318
(2008).

\bibitem{cd11JPA} Lin Chen and D.\v{Z}. {\Dbar}okovi{\'c},
Distillability and PPT entanglement of low rank quantum states,
\jpa: Math. Theor. {\bf44}, 285303 (2011), (26pp).

\bibitem{cd11JMP} Lin Chen and D.\v{Z}. {\Dbar}okovi{\'c},
Description of rank four entangled states of two qutrits having
positive partial transpose, J. Math. Phys. {\bf 52},
122203 (2011), (27pp).

\bibitem{cd11JMPe}  Lin Chen and D.\v{Z}. {\Dbar}okovi{\'c},
Erratum: "Description of rank four entangled states of two qutrits having positive partial transpose" [J. Math. Phys. 52, 122203 (2011)],
J. Math. Phys. {\bf 53}, 079903 (2012).

\bibitem{cd12} Lin Chen and D.\v{Z}. {\Dbar}okovi{\'c}, Properties and construction of extreme bipartite states having positive partial
transpose, quant-ph/1203.1364 (2012).

\bibitem{choi82} M.D. Choi, Positive linear maps, in ``Operator Algebras and Applications'', Kingston, Proc. Sympos. Pure. Math. {\bf38}, Amer. Math. Soc. Part 2, 583-590 (1980).

\bibitem{clarisse06} L. Clarisse,
Construction of bound entangled states with special ranks, Phys.
Lett. A {\bf359} (2006), 603-607.


\bibitem{DiV03}
D.P. DiVincenzo, T. Mor, P.W. Shor, J.A. Smolin, and B.M. Terhal,
Unextendible product bases, uncompletable product bases and bound
entanglement, Commun. Math. Phys. {\bf238} (2003), 379-410.

\bibitem{djokovic11} D. \v{Z} {\Dbar}okovi{\'c},
The checkerboard family of entangled states of two qutrits,
Centr. Eur. J. Phys. {\bf 9} (1) (2011), 65-70.

\bibitem{dvc00} W. D\"{u}r, G. Vidal, and J.I. Cirac, Three qubits can be entangled in two inequivalent ways,
\pra {\bf62}, 062314 (2000).

\bibitem{ekert91} A.K. Ekert,
Quantum cryptography based on Bell's theorem, \prl {\bf67}, 661
(1991).

% \bibitem{Gib}
% C.G. Gibson, Elementary Geometry of Algebraic Curves,
% An Undergraduate Introduction, Cambridge Univ. Press, 1998.

% \bibitem{GP}
% G.-M. Greuel and G. Pfister, A Singular Introduction to Commutative
% Algebra, Springer, New York, 2002.

% \bibitem{ghb02} O. G\"{u}hne, P. Hyllus, D. Bruss, A. Ekert,
% M. Lewenstein, C. Macchiavello, and A. Sanpera,
% Detection of entanglement with few local measurements,
% \pra {\bf66}, 062305 (2002).

% \bibitem{ghg07} O. G\"{u}hne, P. Hyllus, O. Gittsovich, and
% J. Eisert, Covariance matrices and the separability problem,
% \prl {\bf99}, 130504 (2007).

\bibitem{gurvits03} L. Gurvits, Classical deterministic complexity of Edmonds' Problem and quantum entanglement,
in Proceedings of the Thirty-fifth annual ACM Symposium on Theory of
Computing, San Diego, California, June 9-11, 2003, ACM Press, New
York, p10 (2003).

\bibitem{hkp03} K.-C. Ha, S.-H. Kye, and Y.S. Park,
Entanglements with positive partial transposes arising from
indecomposable positive linear maps, \pla~{\bf313}, 163-174 (2003).

\bibitem{ha09} K.-C. Ha, Comment on "Extreme rays in $3\ox3$
entangled edge states with positive partial transposes", \pla
{\bf373}, 2298 (2009).

\bibitem{hk05} K.-C. Ha, and S.-H. Kye, Construction of $3\ox3$
entangled edge states with positive partial transposes, \jpa: Math.
Gen. {\bf38} (2005), 9039-9050.

\bibitem{hhm11} L.O. Hansen, A. Hauge, J. Myrheim, and P.O. Sollid,
Low-rank positive-partial-transpose states and their relation to
product vectors, \pra {\bf85}, 022309 (2012).

%\bibitem{h92} J. Harris, Algebraic Geometry, A First Course,
%Springer-Verlag, New York (1992).

\bibitem{hhh05} K. Horodecki, M. Horodecki, P. Horodecki, and
J. Oppenheim, Secure key from bound entanglement, \prl {\bf94},
160502 (2005).

\bibitem{hhh96} M. Horodecki, P. Horodecki, and R. Horodecki,
Separability of mixed states: necessary and sufficient conditions,
Phys. Lett. A {\bf223} (1996), 1-8.

\bibitem{horodecki97} P. Horodecki, Separability criterion and
inseparable mixed states with positive partial transpose, Phys.
Lett. A {\bf232} (1997), 333-339.

\bibitem{hlv00} P. Horodecki, M. Lewenstein, G. Vidal, and I. Cirac,
Operational criterion and constructive checks for separability of
low rank density matrices, Phys. Rev. A {\bf62}, 032310 (2000).

\bibitem{hst99} P. Horodecki, J.A. Smolin, B.M. Terhal,
A.V. Thapliyal, Rank two bipartite bound entangled states do not
exist, Theoretical Computer Science {\bf 292} (2003), 589-596.

\bibitem{kk07} W.C. Kim and S.H. Kye, Extreme rays in $3\ox3$ entangled edge states with positive partial transposes,
\pla {\bf369}, 16 (2007).

\bibitem{kkl11} Y.H. Kiem, S.H. Kye, and J. Lee, Existence of
product vectors and their partial conjugates in a pair of spaces,
\jmp~{\bf52}, 122201 (2011).

%\bibitem{kye12} S.-H. Kye, Facial structures for various notions of positivity and applications to the theory of entanglement,
%quant-ph/1202.4255 (2012).

\bibitem{ko12} S.H. Kye and H. Osaka, Classification of bi-qutrit
PPT entangled states by their ranks, quant-ph/1202.1699 (2012).

\bibitem{lmo07} J.M. Leinaas, J. Myrheim and E. Ovrum,
Extreme points of the set of density matrices with positive partial
transpose, \pra {\bf76}, 034304 (2007).

%\bibitem{lms10} J.M. Leinaas, J. Myrheim, and P. Sollid, Numerical
%studies of entangled PPT states in composite quantum systems, \pra
%{\bf81}, 062329 (2010).


\bibitem{nc00}  M.A. Nielsen and I.L. Chuang, {\it Quantum
Computation and Quantum Information}
(Cambridge University Press, Cambridge, England, 2000).

\bibitem{peres96} A. Peres, Separability criterion for density
matrices, Phys. Rev. Lett. {\bf77} (1996), 1413-1415.

\bibitem{sbl01} A. Sanpera, D. Bru{\ss}, and M.
Lewenstein, Schmidt number witnesses and bound entanglement,
\pra {\bf63}, 050301 (2001).

\bibitem{skowronek11} L. Skowronek, Three-by-three bound
entanglement with general unextendible product bases, J. Math. Phys. {\bf 52}, 122202 (2011), (32pp).

\bibitem{stormer82} E. St{\o}rmer, Decomposable positive maps
on $\bC^*$-algebras,
Proc. Amer. Math. Soc. {\bf86}, 402 (1982).

\bibitem{werner89} R.F. Werner, Quantum states with
einstein-podolsky-rosen correlations admitting a hidden-variable
model, Phys. Rev. A {\bf40}, 4277 (1989).


\end{thebibliography}
\end{document}